# Explicitly Correlated Double-Hybrid DFT: A Comprehensive Analysis of the Basis Set Convergence on the GMTKN55 Database

Nisha Mehta and Jan M. L. Martin*

**ABSTRACT:** Double-hybrid density functional theory (DHDFT) offers a pathway to accuracy approaching composite wavefunction approaches such as G4 theory. However, the Görling–Levy second-order perturbation theory (GLPT2) term causes them to partially inherit the slow $\propto L^{-3}$ (with $L$ the maximum angular momentum) basis set convergence of correlated wavefunction methods. This could potentially be remedied by introducing F12 explicit correlation: we investigate the basis set convergence of both DHDFT and DHDFT-F12 (where GLPT2 is replaced by GLPT2-F12) for the large and chemically diverse general main-group thermochemistry, kinetics, and noncovalent interactions (GMTKN55) benchmark suite. The B2GP-PLYP-D3(BJ) and revDSD-PBEP86-D4 DHDFs are investigated as test cases, together with orbital basis sets as large as aug-cc-pV5Z and F12 basis sets as large as cc-pVQZ-F12. We show that F12 greatly accelerates basis set convergence of DHDFs, to the point that even the modest cc-pVDZ-F12 basis set is closer to the basis set limit than cc-pV(Q+d)Z or def2-QZVPPD in orbital-based approaches, and in fact comparable in quality to cc-pV(5+d)Z. Somewhat surprisingly, aug-cc-pVDZ-F12 is not required even for the anionic subsets. In conclusion, DHDF-F12/VDZ-F12 eliminates concerns about basis set convergence in both the development and applications of double-hybrid functionals. Mass storage and I/O bottlenecks for larger systems can be circumvented by localized pair natural orbital approximations, which also exhibit much gentler system size scaling.

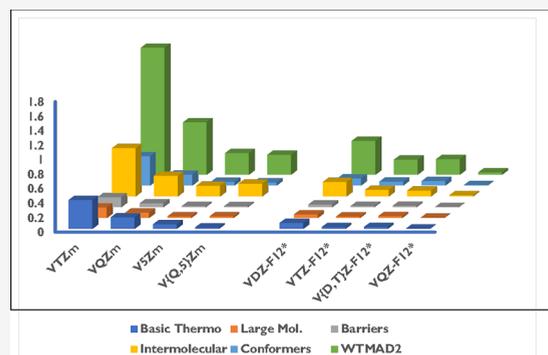

## 1. INTRODUCTION

The two most common methodologies in computational chemistry are wavefunction ab initio methods[1] and density functional theory (DFT).[2,3] Although (correlated) wavefunction ab initio methods provide a clear road map for the convergence to the exact solution, they suffer from slow basis set convergence and hence they are only practical for small molecules. The alternative solution to the quantum many problems is given by DFT, thanks to Hohenberg–Kohn[2] and Kohn–Sham theorems,[3] DFT currently provides the best cost-accuracy ratio for main-group thermochemistry, kinetics, and noncovalent interactions. Among various density functional theory approximations, double-hybrid density functionals (DHDFs) stand out for their general applicability, reliability, and robustness.[4−13] In DHDFs, a portion of semilocal DFT exchange and correlation are replaced by non-local Fock exchange and second-order Görling–Levy perturbation theory[14] (GLPT2) type correlation contributions, respectively. (An earlier usage[15,16] of the term "double hybrid" referred to the combination of semilocal DFT for short-range correlation with regular MP2 correlation in a HF orbital basis for long-range correlation; see also the work of the late Angyán[17] on range-separated correlation. For a detailed numerical analysis of the benefits of GLPT2 over HF-MP2 correlation, see ref 18). DHDFs offer[8,19] a level of agreement approaching composite wavefunction theory schemes such as G3 and G4 theories.[20−22]

Hybrid DFT functionals (rung four on "Jacob's Ladder"[23]) exhibit basis set convergence resembling that of Hartree–Fock theory. Double hybrids (rung five on "Jacob's Ladder") contain a GLPT2 part, the basis set convergence of which is similar to the well-known asymptotic $\propto L^{-3}$ (with $L$ the highest angular momentum in the basis set) behavior of MP2[24] and of electron correlation methods more broadly.[25]

Thus, double hybrids inherit the slow basis set convergence of MP2, although the problem is not as severe as in MP2 itself owing to the scale factors of the GLPT2 correlation (e.g., 0.25 for B2PLYP,[4] 0.36 for B2GP-PLYP[26]). Additionally, the computational cost can be greatly mitigated by introducing density fitting in the MP2 part,[27,28] and two-point basis set extrapolation (e.g., refs 29−31 and references therein) can be applied.

A

The greatest stumbling block for basis set convergence in MP2 and GLPT2 alike is the need to model the interelectronic correlation cusp, which explicitly depends on $r_{12}$, in terms of products of orbitals in $r_1$ and $r_2$. In explicitly correlated approaches (see refs 32−34 for reviews), functions of $r_{12}$ (so-called geminals) are added to the calculation to ensure that the cusp is well-described at short range, "freeing up" the orbital basis set, as it were, to cover other correlation effects.

Kutzelnigg and Morgan[25] showed that for two-electron model systems, the singlet-coupled pair correlation energy converges as $\propto L^{-7}$, compared to $\propto L^{-3}$ for pure orbital calculations.

Initial studies (e.g., refs 35 and 36) featured a simple R12 geminal. In the last decade and a half, the F12 geminal[37] $(1 - \exp \gamma r_{12})/\gamma$ has become the de facto standard. Meanwhile, the computational cost barrier resulting from the need for three- and four-electron integrals[38−40] was circumvented through the introduction of auxiliary basis sets and density fitting.[41−43]

Meanwhile, MP2-F12 and various approximations to CCSD(T)-F12 have become a mainstream tool in high-accuracy wavefunction methods: see, for example, refs 44−46 and from the Weizmann group, refs 47−49 in small-molecule thermochemistry, and refs 50−53 in noncovalent interactions.

It stands to reason that MP2-F12 in a basis of Kohn−Sham orbitals might be a way through the basis set convergence bottleneck of double-hybrid DFT. Karton and Martin[54] showed that this might be the case for a rather small set of closed-shell reactions, but to our knowledge, this has never been verified for a large and chemically diverse benchmark suite such as GMTKN55 (general main-group thermochemistry, kinetics, and noncovalent interactions,[6] 55 problem types) or the Head-Gordon group's even larger main-group chemistry database (MGCDB84[55]). GMTKN55 has previously been used for both evaluation and parametrization of DHDFs as well as composite wavefunction methods.[6,8−10,56−58]

We will show below that for DHDFs applied to GMTKN55, F12 accelerates basis set convergence to the point that even spd basis sets are quite close to the complete basis set limit, and that spdf basis sets effectively reach it.

## 2. COMPUTATIONAL DETAILS

We assess the basis set convergence of conventional and explicitly correlated DHDFs using the GMTKN55 database for general main-group thermochemistry, kinetics, and non-covalent interactions. GMTKN55 consists of 2462 total single point calculations, which are distributed over 55 subsets. The latter are divided into five categories. The first category (basic properties and reaction energies of small systems) addresses problems associated with reaction energies for small systems, total atomization energies, ionization potentials, electron affinities, and self-interaction error. The second category covers problems related to reaction energies of large systems and isomerization. The third category consists of barrier height-related problems. Intermolecular noncovalent interactions related problems are covered in the third category, while conformer equilibria (driven by intramolecular noncovalent interactions) make up the fourth category. The respective abbreviations for the five categories are "Thermo", "Large", "Barrier", "Intermol", and "Conf". Table 1 provides a summary of GMTKN55.

We used the weighted total mean absolute deviation, type 2 (WTMAD2)—originally defined in eq 2 of ref 6—as our primary metric.

Table 1. Overview of the GMTKN55 Database and Its Five Categories: Basic Properties and Reactions of Small Systems ("Thermo"), Reaction Energies of Larger Systems and Isomerization ("Large"), Barrier Heights ("Barrier"), Intermolecular Noncovalent Interactions ("Intermol"), and Intramolecular Noncovalent Interactions ("Conf")[a]

| Category | names of constituent benchmark sets | references |
|---|---|---|
| Thermo | W4-11, G21EA, G21IP, DIPCS10, PA26, SIE4x4, ALKBDE10, YBDE18, AL2X6, HEAVYSB11, NBPRC, ALK8, RC21, G2RC, BH76RC, FH51, TAUT15, DC13 | 6,59−82 |
| Large | MB16-43, DARC, RSE43, BSR36, CDIE20, ISO34, ISOL24, C60ISO, PArel | 6,65,83−89 |
| Barrier | BH76, BHPERI, BHDIV10, INV24, BHROT27, PX13, WCPT18 | 6,68,69,82,90−96 |
| Intermol | RG18, ADIM6, S22, S66, HEAVY28, WATER27, CARBHB12, PNICO23, HAL59, AHB21, CHB6, IL16 | 97−106 |
| Conf | IDISP, ICONF, ACONF, AMINO20x4, PCONF21, MCONF, SCONF, UPU23, BUT14DIOL | 6,67,82,87,107−116 |

[a]For more details, see ref 6.

$$\mathrm{WTMAD2} = \frac{1}{\sum_{i=1}^{55} N_i} \sum_{i=1}^{55} N_i \frac{56.85\ \mathrm{kcal/mol}}{|\overline{\Delta E_i}|} \mathrm{MAD}_i \quad (1)$$

where $N_i$ represents the number of systems in each subset, $|\overline{\Delta E_i}|$ is the mean absolute value of all the reference energies for $i = 1$ to 55, and $\mathrm{MAD}_i$ is the mean absolute deviations of the calculated and reference energies for each subset of GMTKN55.

All electronic structure calculations were performed using the MOLPRO2022 package[117] on the ChemFarm HPC cluster of the Faculty of Chemistry at the Weizmann Institute of Science. The B2GP-PLYP[26]-D3(BJ)[118] and revDSD-PBEP86-D4[8] double hybrids were investigated as test cases. The dispersion model for B2GP-PLYP considered here was DFT-D3 of Grimme et al.[97] with the Becke−Johnson damping function.[119] We used the B2GP-PLYP-D3(BJ)[26] dispersion parametrization $s_6 = 0.560$, $s_8 = 0.2597$, $a_1 = 0.000$, and $a_2 = 6.3332$ from ref 118. For revDSD-PBEP86,[8] we used the DFT-D4 dispersion correction of Grimme et al.[120,121] with the parameters $s_6 = 0.5132$, $s_8 = 0.000$, $a_1 = 0.4400$, $a_2 = 3.60$, and $s_9 = 0.5132$ from ref 8. As per the DFT-D4 defaults, we used electronegativity equalization[122] partial charges and the 3-body Axilrod−Teller−Muto correction term. DFT-D3 and DFT-D4 type dispersion corrections were obtained with the respective standalone programs by Grimme and co-workers.[123,124]

Whenever possible, all of the KS, MP2, and MP2-F12 steps were carried out with density fitting (DF-KS, DF-MP2, and DF-MP2-F12 approximations). We employed the OptRI auxiliary basis set[125] within the complementary auxiliary basis set approach,[126] the JKFIT basis sets of Weigend[127] for the DF-KS calculations, and the MP2FIT set of Hättig and co-workers[128,129] for the DF-MP2/DF-MP2-F12 steps. Throughout the manuscript, DHDF-F12 refers to the double-hybrid calculations with the MP2-F12 (or DF-MP2-F12) method, whereas DHDF refers to the orbital-only (i.e., non-F12) double-hybrid calculations. In all of the DHDF-F12 calculations, the default fixed-amplitude "3C(FIX)" approximation was employed. All self-consistent-field energies were corrected with the complementary auxiliary basis set (CABS) singles correction. Energy convergence criteria for the KS calculations were set to $10^{-9}E_h$ throughout, with MOLPRO's default integration grids for this accuracy and the basis set at hand.

Table 2. Statistical Analysis of the Basis Set Convergence in Conventional and Explicitly Correlated B2GP-PLYP-D3(BJ)) Calculations for the GMTKN55 Database and Its Categories, Relative to the Reference 6 Reference Data[a]

| | B2GP-PLYP-D3(BJ) | | | | | | B2GP-PLYP-F12-D3(BJ) | | | | | |
|---|---|---|---|---|---|---|---|---|---|---|---|---|
| | WTMAD2 | THERMO | BARRIERS | LARGE | CONF | INTERMOL | | WTMAD2 | THERMO | BARRIERS | LARGE | CONF | INTERMOL |
| VDZ | 11.904 | 2.205 | 0.964 | 1.049 | 4.160 | 3.526 | AVDZ-F12 | 3.011 | 0.581 | 0.333 | 0.680 | 0.623 | 0.793 |
| VDZ* | 9.661 | 1.323 | 0.627 | 1.049 | 4.160 | 2.503 | VDZ-F12 | 2.953 | 0.585 | 0.334 | 0.660 | 0.619 | 0.756 |
| VDZ[m] | 6.332 | 1.323 | 0.627 | 1.002 | 1.498 | 1.883 | VDZ-F12* | 2.939 | 0.580 | 0.334 | 0.660 | 0.619 | 0.747 |
| VTZ | 5.649 | 1.062 | 0.553 | 0.698 | 1.405 | 1.930 | VTZ-F12 | 2.979 | 0.584 | 0.331 | 0.652 | 0.587 | 0.825 |
| VTZ* | 4.495 | 0.646 | 0.389 | 0.698 | 1.405 | 1.356 | VTZ-F12* | 2.969 | 0.582 | 0.331 | 0.652 | 0.587 | 0.817 |
| VTZ[m] | 3.427 | 0.646 | 0.389 | 0.694 | 0.634 | 1.064 | V{D,T}Z-F12 | 3.005 | 0.582 | 0.334 | 0.645 | 0.585 | 0.860 |
| VQZ | 3.978 | 0.761 | 0.445 | 0.639 | 0.760 | 1.374 | V{D,T}Z-F12* | 2.993 | 0.581 | 0.333 | 0.645 | 0.585 | 0.849 |
| VQZ* | 3.417 | 0.558 | 0.348 | 0.639 | 0.760 | 1.113 | VQZ-F12 | 3.007 | 0.591 | 0.330 | 0.667 | 0.585 | 0.833 |
| VQZ[m] | 3.131 | 0.558 | 0.348 | 0.646 | 0.590 | 0.990 | VQZ-F12* | 3.004 | 0.589 | 0.327 | 0.666 | 0.584 | 0.838 |
| V{T,Q}Z | 3.955 | 0.738 | 0.448 | 0.625 | 0.673 | 1.472 | V{T,Q}Z-F12 | 3.015 | 0.592 | 0.330 | 0.669 | 0.585 | 0.839 |
| V{T,Q}Z* | 3.521 | 0.593 | 0.353 | 0.625 | 0.673 | 1.277 | V{T,Q}Z-F12* | 3.016 | 0.591 | 0.327 | 0.668 | 0.583 | 0.847 |
| V{T,Q}Z[m] | 3.351 | 0.593 | 0.353 | 0.629 | 0.597 | 1.179 | | | | | | | |
| V5Z* | 3.054 | 0.573 | 0.328 | 0.660 | 0.609 | 0.885 | | | | | | | |
| V5Z[m] | 3.020 | 0.573 | 0.328 | 0.661 | 0.584 | 0.874 | | | | | | | |
| V{Q,5}Z* | 3.105 | 0.589 | 0.326 | 0.668 | 0.593 | 0.930 | | | | | | | |
| V{Q,5}Z[m] | 3.115 | 0.589 | 0.326 | 0.666 | 0.597 | 0.937 | | | | | | | |
| def2-TZVPP | 3.966 | 0.834 | 0.443 | 0.633 | 0.890 | 1.166 | | | | | | | |
| def2-TZVPP* | 3.412 | 0.660 | 0.342 | 0.633 | 0.890 | 0.888 | | | | | | | |
| def2-TZVPP[m] | 3.162 | 0.660 | 0.342 | 0.639 | 0.689 | 0.833 | | | | | | | |
| def2-TZVPPD | 3.157 | 0.657 | 0.333 | 0.582 | 0.685 | 0.900 | | | | | | | |
| def2-QZVPP | 3.267 | 0.653 | 0.364 | 0.643 | 0.624 | 0.984 | | | | | | | |
| def2-QZVPP* | 3.007 | 0.591 | 0.322 | 0.643 | 0.624 | 0.828 | | | | | | | |
| def2-QZVPP[m] | 2.953 | 0.591 | 0.322 | 0.645 | 0.592 | 0.803 | | | | | | | |
| def2-QZVPPD | 2.965 | 0.595 | 0.318 | 0.630 | 0.584 | 0.837 | | | | | | | |
| def2-{T,Q}ZVPP | 3.326 | 0.651 | 0.371 | 0.657 | 0.598 | 1.050 | | | | | | | |
| def2-{T,Q}ZVPP* | 3.177 | 0.623 | 0.329 | 0.657 | 0.598 | 0.969 | | | | | | | |
| def2-{T,Q}ZVPP[m] | 3.187 | 0.623 | 0.329 | 0.657 | 0.612 | 0.965 | | | | | | | |
| def2-{T,Q}ZVPPD | 3.111 | 0.625 | 0.321 | 0.670 | 0.600 | 0.894 | | | | | | | |
| VDZ-F12 | 5.883 | 0.880 | 0.323 | 0.916 | 1.539 | 2.225 | | | | | | | |

[a]VnZ*: AVnZ was employed for RG18, AHB21, G21EA, IL16, WATER27, BH76, and BH76RC. In the "VnZ[m]" variant, we additionally treated the BUT14DIOL, S22, S66, SCONF, PNICO23, PCONF21, PArel, MCONF, and AMINO20x4 test sets with the hAVnZ basis set. VnZ-F12*: AVnZ-F12 was employed for RG18, AHB21, G21EA, IL16, WATER27, BH76, and BH76RC. In the "VnZ-F12[m]" variant, we additionally treated the BUT14DIOL, S22, S66, SCONF, PNICO23, PCONF21, PArel, MCONF, and AMINO20x4 test sets with the hAVnZ-F12 basis set.

We considered different families of basis sets. The first category are the correlation consistent basis sets of Dunning,[130−132] which were developed with orbital-based correlated wavefunction calculations in mind (optimized for CISD valence correlation energies of atoms). The notation VnZ, in this paper, is shorthand for the combination of regular cc-pVnZ on first-row elements, cc-pV(n+d)Z on second-row elements, and cc-pVnZ-PP for the heavy p-block elements, where PP stands for pseudopotential. Finally, we employed ad hoc modifications: for RG18[6] and the anion-containing subsets AHB21,[106] G21EA,[60,82] IL16,[106] WATER27,[101,102] BH76,[6,68,70,82] and BH76RC,[6,82] we employed aug-cc-pVnZ ("VnZ*"). In the "VnZ$^m$" variant, we additionally treated the BUT14DIOL,[6,116] S22,[98,99] S66,[100] SCONF,[6,82,114] PNICO23,[6,103] PCONF21,[6,111,112] PArel,[6] MCONF,[6,133] and AMINO20x4[6,110] test sets with the hAVnZ basis set (cc-pVnZ on hydrogen, aug-cc-pVnZ on first-row elements, aug-cc-pV(n+d)Z on second-row elements, and aug-cc-pVnZ-PP for the heavy p-block elements).

The second class of basis sets considered are the cc-pVnZ-F12 (abbreviated VnZ-F12 in this manuscript) of Peterson and co-workers,[134] or their anionic-friendly variants aug-cc-pVnZ-F12 (AVnZ-F12).[135] These basis sets were explicitly developed with F12 calculations in mind. In fact, non-F12 basis sets in explicitly correlated calculations lead to non-monotonous convergence because of elevated and erratic basis set superposition errors (BSSEs).[49] VnZ-F12* indicates that the VnZ-F12 basis set was used for all subsets of GMTKN55 except WATER27, IL16, G21EA, BH76, BH76RC, AHB21, and RG18, where we used AVnZ-F12. Again, we employed cc-pVnZ-F12-PP for the heavy p-block elements. The geminal Slater exponent ($\beta$) values of 0.9, 1.0, and 1.0 were used for the (A-)VDZ-F12, (A-)VTZ-F12, and (A-)VQZ-F12, respectively.

Finally, we also considered the Weigend−Ahlrichs/Karlsruhe def2 family,[136] namely def2-TZVPP and def2-QZVPP, and their diffuse function-augmented variants def2-TZVPPD and def2-QZVPPD.[137] def2-nZVPP* and def2-nZVPP$^m$ variants are defined analogously to the above.

The geometries, charge/multiplicity information and reference energies were obtained from ref 6 and used verbatim throughout. The most computationally demanding subset isomerization energies of fullerene $C_{60}$ structures (C60ISO)[89] might just barely have been feasible with the VDZ-F12 basis set with available computational resources, but near-singularity in the overlap matrix (smallest eigenvalue $3 \times 10^{-11}$) effectively made the KS calculations impossible to converge. This subset's omission does not significantly affect WTMAD2 because of its small weight in the WTMAD2 formula. For explicitly correlated DHDF calculations on the UPU23 subset,[115] we settled for (A-)VDZ-F12 basis to reduce computational cost.

## 3. RESULTS AND DISCUSSION

### 3.1. Basis Set Convergence for Conventional Double Hybrids. 
Let us first consider the basis set convergence with the orbital basis sets in conventional double-hybrid calculations, that is, B2GP-PLYP-D3(BJ)/VnZ, where n = D,T,Q, and 5 (Table 2). The PT2 component slows down basis set convergence, albeit mitigated (compared to MP2 in a Hartree−Fock basis set) by the PT2 coefficients in the double hybrid (typically in the 0.1−0.5 range). Although DHDFs converge faster than ab initio methods, their PT2 part acquires a slower basis set convergence. The VDZ basis set yields an unacceptably large WTMAD2 of 11.904 kcal/mol for the entire GMTKN55 database. This goes down to 9.661 kcal/mol when substituting the AVDZ basis set for the rare gas clusters RG18 and the six anion-containing subsets WATER27, BH76, BH76RC, AHB21, G21EA, and IL16. A further reduction to 6.332 kcal/mol was achieved for the VDZ$^m$ variant, where the haVDZ basis set additionally was applied to BUT14DIOL, S22, S66, SCONF, PNICO23, PCONF21, PArel, MCONF, and AMINO20x4. Therefore, we will mostly discuss our statistics of conventional double-hybrid calculations with the VnZ$^m$ variant. The VTZ$^m$ basis set nearly halves WTMAD2 to 3.427 kcal/mol. In order to surpass this level of accuracy, VQZ$^m$ has to be employed, yielding a WTMAD2 of 3.131 kcal/mol. For still better basis set convergence, we employed V5Z$^m$, which slightly further lowers WTMAD2 to 3.020 kcal/mol. As the orbital-only B2GP-PLYP-D3(BJ) complete basis set limit estimate, we extrapolate VQZ$^m$ and V5Z$^m$ reaction energies using the two-point extrapolation formula ($A + B/L^\alpha$, L = highest angular momentum present in the basis set) where $\alpha$ = 8.7042 for KS and 2.7399 for PT2 (as recommended in refs 138 and 139) components, respectively. The B2GP-PLYP-D3(BJ)/V\{Q,5\}Z$^m$ level of theory results in a WTMAD2 of 3.115 kcal/mol for the entire GMTKN55 database. B2GP-PLYP-D3(BJ)/V\{T,Q\}Z$^m$ ($\alpha$ = 7.6070 for KS and 2.5313 for PT2) yields WTMAD2 of 3.351 kcal/mol.

A breakdown into the five top-level subdivisions of GMTKN55 (Table 2) showed that all five of them smoothly approach the basis set limit at the B2GP-PLYP-D3(BJ)/V\{Q,5\}Z$^m$ level. More detailed scrutiny of the individual subsets revealed that HEAVY28[6,97] is the major contributor to the difference between V5Z$^m$ and V\{Q,5\}Z$^m$, with $\Delta$WTMAD2 increased by 0.048 kcal/mol. Because of the way HEAVY28, RG18, and HAL59[6,104,105] are weighted in WTMAD2, a small change in those subsets has an outsize contribution.

### 3.2. Effect of Introducing F12 Terms. 
Next, we investigate the basis set convergence in the explicitly correlated double-hybrid calculations. These calculations need to be done with the cc-pVnZ-F12 basis sets of Peterson et al.[134] Table 2 presents a statistical analysis of B2GP-PLYP-F12-D3(BJ) calculations. The B2GP-PLYP-F12-D3(BJ)/VDZ-F12 level of theory results in a WTMAD2 of only 2.953 kcal/mol. We would like to emphasize that this is *not* just a matter of the basis set: for illustration, we also evaluated orbital-only B2GP-PLYP-D3(BJ)/VDZ-F12 and found that WTMAD2 shot up to 5.883 kcal/mol. The difference is entirely owing to the presence versus absence of geminal F12 terms in the GLPT2 evaluation. Somewhat surprisingly, WTMAD2 with VDZ-F12* basis (AVDZ-F12 basis for the rare gas clusters RG18 and six anion containing subsets WATER27, BH76, BH76RC, AHB21, G21EA, and IL16) only was reduced to 2.939 kcal/mol, indicating that not even for anionic subsets is AVDZ-F12 required. (We do note that, unlike the VDZ orbital basis set, the VDZ-F12 already includes one diffuse function each of s and p symmetries for p-block elements, and one diffuse s function for hydrogen.) In explicitly correlated B2GP-PLYP-F12-D3(BJ), the energy differences that make up the GMTKN55 benchmark converge markedly, one might even say dramatically, faster with respect to the basis set size. For example, VDZ-F12*, VTZ-F12*, and VQZ-F12* provide WTMAD2 which are 2.939, 2.969, and 3.004 kcal/mol above the reference values, respectively. Small discrepancies between the three basis sets are mostly because of rare-gas clusters RG18 with their outsize weights, which contribute 0.042 and 0.016 kcal/mol, respectively, toward the increase in WTMAD2 for VDZ-F12* to VTZ-F12* and VTZ-F12* to VQZ-F12*. The other test sets that contribute toward deviations between VDZ-F12* and VTZ-F12* are HEAVY28 (0.021 kcal/mol) and HAL59 (0.008 kcal/mol).

Table 3. A Comparison of Total WTMAD2 of GMTKN55 Data Set (i.e., WTMAD2 (all)) and WTMAD2 after Excluding RG18, HEAVY28, and HAL59 from the Statistics (i.e., WTMAD2 (mod))[a]

| | WTMAD2 (all) | WTMAD2 (mod.) | WTMAD2 (all) | WTMAD2 (mod.) |
|---|---|---|---|---|
| | GMTKN55 as reference | | CBS limit as reference | |
| | B2GP-PLYP-F12-D3(BJ) | | | |
| AVDZ-F12 | 3.011 | 2.706 | 0.418 | 0.320 |
| VDZ-F12 | 2.953 | 2.686 | 0.499 | 0.391 |
| VDZ-F12* | 2.939 | 2.671 | 0.467 | 0.358 |
| VTZ-F12 | 2.979 | 2.632 | 0.220 | 0.191 |
| VTZ-F12* | 2.969 | 2.626 | 0.207 | 0.172 |
| V{D,T}Z-F12 | 3.005 | 2.625 | 0.232 | 0.205 |
| V{D,T}Z-F12* | 2.993 | 2.621 | 0.215 | 0.191 |
| VQZ-F12 | 3.007 | 2.657 | 0.065 | 0.042 |
| VQZ-F12* | 3.004 | 2.649 | 0.032 | 0.024 |
| V{T,Q}Z-F12 | 3.015 | 2.662 | 0.038 | 0.018 |
| V{T,Q}Z-F12* | 3.016 | 2.653 | 0.002 | 0.001 |
| | B2GP-PLYP-D3(BJ) | | | |
| VDZ | 11.904 | 11.295 | 11.303 | 10.825 |
| VDZ* | 9.661 | 9.120 | 9.014 | 8.582 |
| VDZ$^m$ | 6.332 | 5.540 | 5.602 | 4.911 |
| VTZ | 5.649 | 5.185 | 4.317 | 3.942 |
| VTZ* | 4.495 | 4.132 | 3.020 | 2.779 |
| VTZ$^m$ | 3.427 | 2.984 | 1.752 | 1.414 |
| VQZ | 3.978 | 3.517 | 1.913 | 1.717 |
| VQZ* | 3.417 | 2.969 | 1.172 | 1.065 |
| VQZ$^m$ | 3.131 | 2.662 | 0.723 | 0.582 |
| V{T,Q}Z | 3.955 | 3.334 | 1.578 | 1.264 |
| V{T,Q}Z* | 3.521 | 2.892 | 0.977 | 0.733 |
| V{T,Q}Z$^m$ | 3.351 | 2.710 | 0.596 | 0.323 |
| V5Z* | 3.054 | 2.678 | 0.372 | 0.278 |
| V5Z$^m$ | 3.020 | 2.641 | 0.299 | 0.199 |
| V{Q,5}Z* | 3.105 | 2.660 | 0.302 | 0.167 |
| V{Q,5}Z$^m$ | 3.115 | 2.671 | 0.275 | 0.138 |
| def2-TZVPP | 3.966 | 3.701 | 2.534 | 2.369 |
| def2-TZVPP* | 3.412 | 3.158 | 1.883 | 1.750 |
| def2-TZVPP$^m$ | 3.162 | 2.889 | 1.633 | 1.480 |
| def2-TZVPPD | 3.157 | 2.781 | 1.530 | 1.405 |
| def2-QZVPP | 3.267 | 2.878 | 1.045 | 0.899 |
| def2-QZVPP* | 3.007 | 2.671 | 0.748 | 0.636 |
| def2-QZVPP$^m$ | 2.953 | 2.613 | 0.750 | 0.638 |
| def2-QZVPPD | 2.965 | 2.577 | 0.743 | 0.625 |
| def2-{T,Q}ZVPP | 3.326 | 2.856 | 0.743 | 0.560 |
| def2-{T,Q}ZVPP* | 3.177 | 2.719 | 0.573 | 0.387 |
| def2-{T,Q}ZVPP$^m$ | 3.187 | 2.730 | 0.519 | 0.329 |
| def2-{T,Q}ZVPPD | 3.111 | 2.724 | 0.456 | 0.289 |

[a]All values are reported in kcal/mol.

A reviewer inquired whether the mildly non-monotonic basis set convergence observed in Table 2 could be attributable to the excessive weights given to the three subsets RG18, HEAVY28, and HAL59 in eq 1. Table 3 compares convergence behavior for GMTKN55 including versus excluding these three subsets from the summations in the numerator and denominator of eq 1. Although the "exclusive" WTMAD2 values are naturally considerably smaller, the mild non-monotonicity persists and likely needs to be attributed to subtle error compensations in the individual subsets between basis set incompleteness and intrinsic functional error. Consistent with this conjecture, we observe that WTMAD2 values relative to the complete basis set limit of the functional (right-hand pane of Table 3) do converge monotonically both with and without the said three subsets.

WTMAD2 obtained with V{D,T}Z-F12* (2.993 kcal/mol) and V{T,Q}Z-F12* (3.016 kcal/mol) pairs can essentially be regarded as the basis set limit. We used the two-point extrapolation formula $(A + B/L^\alpha$, $L =$ highest angular momentum present in the basis set) for the PT2 components with $\alpha = 3.0878$ for the V{D,T}Z-F12 pair and $\alpha = 4.3548$ for the V{T,Q}Z-F12 pair.[138] The extrapolation of the KS component essentially provides the same WTMAD2 as obtained with just the highest angular momentum present in the basis set and CABS. Switching off the CABS correction only increases the WTMAD2 value for V{D,T}Z-F12 from 3.005 to 3.013 kcal/mol.

In order to explore whether MP2-F12 extrapolation exponents can be safely used for the PT2-F12 component in DHDF-F12, we performed a sensitivity analysis of B2GP-PLYP-F12-D3(BJ)/V{D,T}Z-F12 extrapolation by calculating rmsd differences [rmsd(extrapolation exponent $\alpha$)−rmsd($\infty$)] for the atomization energies of the W4-11 set calculated relative to B2GP-PLYP-F12-D3(BJ)/V{T,Q}Z-F12. Figure 1 shows a minimum near $\alpha$ = 3.4. $\alpha$ = 3.0878 taken from ref 138 yields rmsd($\alpha$)−rmsd($\infty$) = −0.040 kcal/mol instead of −0.041 kcal/mol for $\alpha$ = 3.4, which is a negligible difference in the larger scheme of things. For different double hybrids, the minimum of this shallow curve might vary slightly around $\alpha$ = 3.4, without significantly affecting rmsd. Hence, we elected to retain the MP2-F12 extrapolation exponent.

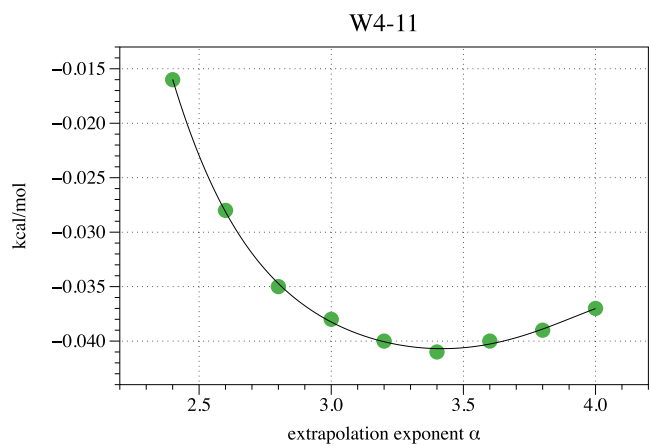

Figure 1. Sensitivity analysis of the B2GP-PLYP-F12-D3(BJ)/V{D,T}Z extrapolation. Root-mean-square deviation (rmsd) differences [rmsd(extrapolation exponent $\alpha$—rmsd($\infty$)] for the atomization energies of the W4-11[59] set calculated relative to B2GP-PLYP-F12-D3(BJ)/V{T,Q}Z-F12.

**3.3. Aside on BSSE.** A brief digression on BSSE might shed more light on basis set convergence behavior. For the intermolecular subset of GMTKN55, one has the option of applying counterpoise (CP) corrections[140] (for detailed discussion and further references, see Burns et al.[141] for WFT methods, Brauer et al.[142] for F12 methods, and ref 143 for DFT and double hybrids). For the intramolecular subset, CP corrections would be rather more awkward, although geometric CP corrections do exist[144,145] for some levels of theory. (For an alternative approach to noncovalent interactions for large systems, involving small tailored basis sets, see ref 146 and references therein). Hence, most groups that employ GMTKN55 avoid CP corrections, which of course presuppose basis sets large enough that these no longer matter (much).

One major benefit of F12 methods (with F12 basis sets) was previously found to be[142,147] a drastic reduction in BSSE, as shown for thermochemistry[147] and for noncovalent interactions.[142,143]

Table 4 presents CP corrections for the Watson−Crick and stacked uracil dimers (systems 17 and 26, respectively, in S66), as representative examples of strong hydrogen bonding and $\pi$-stacking, respectively. As seen in Table 4, B2GP-PLYP-F12/V$n$Z-F12 leads to a BSSE reduction by an order of magnitude (or more) over the corresponding B2GP-PLYP/V$n$Z calculation, and indeed one has to go all the way to V5Z to find a basis set with a similarly low BSSE as B2GP-PLYP-F12/VDZ-F12 (!). For haV$n$Z-F12 versus haV$n$Z, and for AV$n$Z-F12 versus AV$n$Z, one likewise sees one order of magnitude reduction in BSSE. Additionally, AV$n$Z-F12 further reduced BSSE by about a factor 2−3 over the already low values for V$n$Z-F12.

At the CBS limit, the BSSE correction should of course be zero, as raw and CP-corrected calculations should yield the same answer. The deviation from zero when extrapolating CP corrections to the CBS limit is a good proxy for the quality of the extrapolation (and its underlying basis sets). For V{T,Q}Z-F12 and AV{T,Q}Z-F12, this evidently works beautifully. For V{D,T}Z-F12 and AV{D,T}Z-F12, not much improvement over the already low BSSE of VTZ-F12 viz. AVTZ-F12 can be seen. For V{Q,5}Z, on the other hand, we find a large negative BSSE that indicates overcorrection. In fact, simple VDZ-F12 has less BSSE than V{Q,5}Z and similar to haV{Q,5}Z.

**3.4. Basis Set Convergence Relative to the Complete Basis Set Limit.** Furthermore, we explored the basis set convergence of conventional and explicitly correlated double-hybrid calculations using basis set limit reference values (Table 5). For this purpose, we used energies calculated at the B2GP-PLYP-F12-D3(BJ)/V{T,Q}-F12* level of theory, as they are sufficiently converged to the basis set limit. Conventional

Table 4. B2GP-PLYP-F12 Compared to B2GP-PLYP CP Corrections (kcal/mol) for the Two Uracil Dimer Structures in S66 Using Different Basis Sets[a]

| | dimer 17 Watson-Crick | dimer 26 $\pi$-stacked | | dimer 17 Watson-Crick | dimer 26 $\pi$-stacked | | dimer 17 Watson-Crick | dimer 26 $\pi$-stacked |
|---|---|---|---|---|---|---|---|---|
| B2GP-PLYP-F12 | | | | | | | | |
| VDZ-F12 | 0.191 | 0.317 | haVDZ-F12 | 0.107 | 0.128 | AVDZ-F12 | 0.097 | 0.124 |
| VTZ-F12 | 0.106 | 0.208 | haVTZ-F12 | 0.056 | 0.113 | AVTZ-F12 | 0.065 | 0.116 |
| VQZ-F12 | 0.042 | 0.068 | haVQZ-F12 | 0.014 | 0.026 | AVQZ-F12 | 0.017 | 0.027 |
| B2GP-PLYP | | | | | | | | |
| VDZ | 4.768 | 4.545 | haVDZ | 1.285 | 2.927 | AVDZ | 1.848 | 3.083 |
| VTZ | 1.630 | 2.089 | haVTZ | 0.650 | 1.120 | AVTZ | 0.901 | 1.223 |
| VQZ | 0.634 | 0.908 | haVQZ | 0.261 | 0.465 | AVQZ | 0.348 | 0.507 |
| V5Z | 0.210 | 0.289 | haV5Z | 0.108 | 0.169 | AV5Z | 0.160 | 0.188 |

[a]haV$n$Z-F12, by analogy with haV$n$Z, corresponds to AV$n$Z-F12 on nonhydrogen elements and V$n$Z-F12 on hydrogen.

Table 5. Statistical Analysis of the Basis Set Convergence in Conventional and Explicitly Correlated B2GP-PLYP-D3(BJ) Calculations for the GMTKN55 Database and Its Categories, Relative to the B2GP-PLYP-F12-D3(BJ)/V{T,Q}Z-F12* Reference Data[a]

| | B2GP-PLYP-D3(BJ) | | | | | | | B2GP-PLYP-F12-D3(BJ) | | | | | |
|---|---|---|---|---|---|---|---|---|---|---|---|---|---|
| | WTMAD2 | THERMO | BARRIERS | LARGE | CONF | INTERMOL | | WTMAD2 | THERMO | BARRIERS | LARGE | CONF | INTERMOL |
| VDZ | 11.303 | 2.251 | 0.869 | 0.854 | 4.110 | 3.220 | AVDZ-F12 | 0.418 | 0.087 | 0.036 | 0.061 | 0.083 | 0.152 |
| VDZ* | 9.014 | 1.340 | 0.464 | 0.854 | 4.110 | 2.246 | VDZ-F12 | 0.499 | 0.096 | 0.042 | 0.050 | 0.098 | 0.212 |
| VDZ[m] | 5.602 | 1.340 | 0.464 | 0.755 | 1.416 | 1.627 | VDZ-F12* | 0.467 | 0.082 | 0.038 | 0.050 | 0.098 | 0.199 |
| VTZ | 4.317 | 0.853 | 0.373 | 0.174 | 1.325 | 1.592 | VTZ-F12 | 0.220 | 0.035 | 0.019 | 0.021 | 0.055 | 0.090 |
| VTZ* | 3.020 | 0.399 | 0.137 | 0.174 | 1.325 | 0.984 | VTZ-F12* | 0.207 | 0.025 | 0.016 | 0.021 | 0.055 | 0.090 |
| VTZ[m] | 1.752 | 0.399 | 0.137 | 0.145 | 0.404 | 0.666 | V{D,T}Z-F12 | 0.232 | 0.033 | 0.023 | 0.027 | 0.062 | 0.088 |
| VQZ | 1.913 | 0.407 | 0.201 | 0.082 | 0.431 | 0.792 | V{D,T}Z-F12* | 0.215 | 0.026 | 0.020 | 0.027 | 0.062 | 0.079 |
| VQZ* | 1.172 | 0.160 | 0.054 | 0.082 | 0.431 | 0.446 | VQZ-F12 | 0.065 | 0.012 | 0.006 | 0.004 | 0.006 | 0.038 |
| VQZ[m] | 0.723 | 0.160 | 0.054 | 0.075 | 0.151 | 0.284 | VQZ-F12* | 0.032 | 0.005 | 0.002 | 0.004 | 0.006 | 0.016 |
| V{T,Q}Z | 1.578 | 0.244 | 0.180 | 0.075 | 0.246 | 0.833 | V{T,Q}Z-F12 | 0.038 | 0.006 | 0.004 | 0.000 | 0.001 | 0.028 |
| V{T,Q}Z* | 0.977 | 0.061 | 0.045 | 0.075 | 0.246 | 0.550 | V{T,Q}Z-F12* | REF | REF | REF | REF | REF | REF |
| V{T,Q}Z[m] | 0.596 | 0.061 | 0.045 | 0.073 | 0.066 | 0.351 | | | | | | | |
| V5Z* | 0.372 | 0.063 | 0.016 | 0.023 | 0.091 | 0.179 | | | | | | | |
| V5Z[m] | 0.299 | 0.063 | 0.016 | 0.023 | 0.050 | 0.147 | | | | | | | |
| V{Q,5}Z* | 0.302 | 0.021 | 0.014 | 0.027 | 0.063 | 0.178 | | | | | | | |
| V{Q,5}Z[m] | 0.275 | 0.021 | 0.014 | 0.024 | 0.042 | 0.175 | | | | | | | |
| def2-TZVPP | 2.534 | 0.600 | 0.240 | 0.169 | 0.694 | 0.831 | | | | | | | |
| def2-TZVPP* | 1.883 | 0.400 | 0.098 | 0.169 | 0.694 | 0.523 | | | | | | | |
| def2-TZVPP[m] | 1.633 | 0.400 | 0.098 | 0.157 | 0.474 | 0.505 | | | | | | | |
| def2-TZVPPD | 1.530 | 0.326 | 0.074 | 0.175 | 0.473 | 0.483 | | | | | | | |
| def2-QZVPP | 1.045 | 0.253 | 0.106 | 0.082 | 0.202 | 0.402 | | | | | | | |
| def2-QZVPP* | 0.748 | 0.171 | 0.038 | 0.082 | 0.202 | 0.254 | | | | | | | |
| def2-QZVPP[m] | 0.750 | 0.171 | 0.038 | 0.082 | 0.201 | 0.258 | | | | | | | |
| def2-QZVPPD | 0.743 | 0.157 | 0.030 | 0.078 | 0.209 | 0.269 | | | | | | | |
| def2-{T,Q}ZVPP | 0.743 | 0.132 | 0.077 | 0.070 | 0.091 | 0.372 | | | | | | | |
| def2-{T,Q}ZVPP* | 0.573 | 0.089 | 0.025 | 0.070 | 0.091 | 0.298 | | | | | | | |
| def2-{T,Q}ZVPP[m] | 0.519 | 0.089 | 0.025 | 0.067 | 0.085 | 0.254 | | | | | | | |
| def2-{T,Q}ZVPPD | 0.456 | 0.090 | 0.020 | 0.033 | 0.088 | 0.225 | | | | | | | |
| VDZ-F12 | 5.324 | 0.795 | 0.324 | 0.576 | 1.238 | 2.392 | | | | | | | |

[a]Values are heat-mapped from red for the largest via yellow for median to green for the smallest. Note that values are heat-mapped separately for each category of GMTKN55 and the entire database.

B2GP-PLYP-D3(BJ) calculations in conjunction with the VDZ[m] basis set yield a WTMAD2 value that is 5.602 kcal/mol above the basis set limit. Increasing the basis set to VTZ[m] and VQZ[m] reduces this deviation to 1.752 and 0.723 kcal/mol, respectively. V5Z[m] yields a deviation that is only 0.299 kcal/mol above our best estimate (B2GP-PLYP-F12-D3(BJ)/V{T,Q}-F12*). Basis set limit reaction energies for the conventional B2GP-PLYP-D3(BJ)/V{Q,5}[m] calculations differ by only 0.275 kcal/mol from explicitly correlated B2GP-PLYP-F12-D3(BJ)/V{T,Q}-F12*, of which inter- and intramolecular noncovalent interactions account for the lion's share. A closer inspection of the individual subsets revealed that HEAVY28, HAL59, and RG18 are the three largest contributors to the discrepancies, their ΔWTMAD2 of HEAVY28, RG18, and HAL59 being 0.086, 0.034, and 0.027 kcal/mol, respectively, relative to B2GP-PLYP-F12-D3(BJ)/V{T,Q}-F12*. As discussed above, the way these three subsets are weighted in the WTMAD2 formula, a small change in reaction energies has a disproportionate contribution to WTMAD2.

B2GP-PLYP-D3(BJ) in conjunction with def2-TZVPP, def2-QZVPP, and def2-{T,Q}ZVPP ($\alpha$ = 7.6070 for KS and 2.5313 for PT2) basis sets provides WTMAD2 which are 2.534, 1.045, and 0.743 kcal/mol above our best estimate, respectively. Adding diffuse functions to RG18, AHB21, BH76, BH76RC, IL16, G21EA, and WATER27 (i.e., def2-nZVPP* basis set) lowers the WTMAD2 values to 1.883, 0.748, and 0.573 kcal/mol, respectively, for TZ, QZ, and {T,Q}Z basis. On the other hand, def2-TZVPPD, def2-QZVPPD, and def2-{T,Q}ZVPPD provide WTMAD2 which are 1.530, 0.743, and 0.456 kcal/mol, respectively.

Turning our attention to explicitly correlated B2GP-PLYP-F12-D3(BJ) calculations with VnZ-F12 type basis sets, we note that VDZ-F12* already yields an acceptable WTMAD2 which is only 0.467 kcal/mol from the F12 basis set limit. Moving on to AVDZ-F12 provides a WTMAD2 which is just 0.050 kcal/mol below VDZ-F12*. The WTMAD2 component breakdown revealed that S66, HEAVY28, and AMINO20x4 together account for 0.043 kcal/mol of the total improvement in ΔWTMAD2 of AVDZ-F12 in comparison to VDZ-F12*. Increasing the basis set size to VTZ-F12* yields a WTMAD2 of 0.207 kcal/mol. WTMAD2 obtained with the VQZ-F12* basis set (0.032 kcal/mol) can essentially be regarded as the basis set limit.

**3.5. Note on Systematic Errors.** Thus far, we have only discussed WTMAD2. It would be of interest to compare, as a measure of systematic error, the basis set convergence of conventional and explicitly correlated double hybrids in terms of weighted total mean signed deviations (WTMSDs), where MAD in eq 1 is replaced by the MSD. (For clarity, the way the reference data's sign conventions work, a positive WTMSD2 indicates overbinding, and a negative one underbinding). Tables S1 and S2 in Supporting Information present the WTMSD2 of conventional and explicitly correlated B2GP-PLYP-D3(BJ) relative to the reference data[6] and to the complete basis set limit as obtained at the B2GP-PLYP-F12-D3(BJ)/V{T,Q}Z-F12* level. WTMSD2 obtained for B2GP-PLYP-F12-D3(BJ) with VDZ-F12* (0.081 kcal/mol), VTZ-F12* (0.071 kcal/mol), and V{D,T}Z-F12* (0.087 kcal/mol) basis sets are all close to zero and indicate that there is little systematic error relative to the complete basis set limit (Table S2). Turning our attention to conventional B2GP-PLYP-D3(BJ), we obtained WTMSD2 of 2.693, 1.077, 0.501, 0.076, and 0.042 kcal/mol for the VDZ*, VTZ*, VQZ*, V5Z*, and V{Q,5}Z* basis sets, respectively. These results indicate that even VQZ* still overbinds significantly due to BSSE: a breakdown into the five top-level subsets (Table S2) reveals that intermolecular interactions account for about two-thirds, and conformers for almost all the remainder. Turning to the V$nZ^m$ variants, VDZ$^m$ yields a deceptively low WTMSD2 = 0.345 kcal/mol owing to compensation between sizable systematic underestimates of basic thermochemistry, barrier heights, a large molecule reaction, systematic overestimates of conformer energies, and especially intermolecular interactions. Although VTZ$^m$ sees WTMSD2 "only" reduced to 0.235 kcal/mol, the constituent values for the five top-level subsets are actually reduced by factors of 3–4. (This, incidentally, illustrates the dangers of blindly relying on a single global metric). The 0.182 kcal/mol for VQZ$^m$ is mostly driven by the noncovalent interactions (0.213 kcal/mol), slightly compensated by basic thermochemistry and barrier heights. Finally, V5Z$^m$ has only a mild systematic error, again mostly from noncovalent interactions. V{T,Q}Z$^m$ extrapolation yields a WTMSD2 of 0.337 kcal/mol, which at first sight seems inferior even to VDZ$^m$; however, closer inspection reveals that essentially *all* of that number comes from overbinding in the intermolecular interactions due to BSSE and that the remaining four top-level subsets are nearly free of systematic bias. For V{Q,5}Z$^m$, WTMSD2 is down to a paltry 0.048 kcal/mol, essentially all of it again from intermolecular interactions. Interestingly, def2-QZVPP (−0.228 kcal/mol) and especially def2-QZVPP* (−0.024 kcal/mol) and def2-QZVPPD (−0.016 kcal/mol) have much smaller WTMSD2 values, also for the top-level subsets: the negative signs reflect mostly underestimates for small-molecule thermochemistry. The different signs of the def2-{T,Q}ZVPPD* and def2-{T,Q}ZVPPD WTMSD2s essentially reflect systematic overbinding versus underbinding of intermolecular interactions, where the former lacks diffuse functions on such subsets as S22, S66, and the like.

Table 6. Relative CPU Timings for the B2GPPLYP-D3(BJ) and B2GPPLY-F12-D3(BJ) Calculations for $(C_nH_{n+2})_2$[a]

| | B2GP-PLYP-F12 | | B2GP-PLYP | | | | | | |
|---|---|---|---|---|---|---|---|---|---|
| | PNO-LPT2 | canonical | | | | | | | |
| | PNO-F12 VDZ-F12 | VDZ-F12 | {T,Q}ZVPP | {T,Q}ZVPPD | V{T,Q}Z | haV{T,Q}Z | haV{Q,5}Z | AV{T,Q}Z |
| (ethane)$_2$ | 1.08 | 1.00 | 1.38 | 1.92 | 1.74 | 2.02 | 7.20 | 3.69 |
| (propane)$_2$ | 1.05 | 1.00 | 1.22 | 1.84 | 1.53 | 2.19 | 8.33 | 4.15 |
| (butane)$_2$ | 0.83 | 1.00 | 0.91 | 1.47 | 1.58 | 1.84 | 7.15 | 3.47 |
| (pentane)$_2$ | 0.63 | 1.00 | 0.70 | 1.16 | 0.91 | 1.49 | 5.83 | 2.78 |
| (hexane)$_2$ | 0.49 | 1.00 | 0.57 | 0.95 | 0.73 | 1.23 | 4.93 | 2.30 |
| (heptane)$_2$ | 0.38 | 1.00 | 0.48 | 0.81 | 0.62 | 1.05 | 4.24 | 1.96 |
| (n-octane)$_2$ | 0.28 | 1.00 | 0.38 | 0.66 | 0.49 | 0.85 | 3.43 | 1.58 |
| (n-nonane)$_2$ | 0.22 | 1.00 | 0.32 | 0.56 | 0.41 | 0.72 | 2.96 | 1.34 |
| (n-decane)$_2$ | 0.18 | 1.00 | 0.28 | 0.49 | 0.36 | 0.63 | 2.60 | 1.16 |
| (n-dodecane)$_2$ | 0.11 | 1.00 | 0.22 | 0.39 | 0.28 | 0.49 | 2.09 | 0.92 |

[a]Timing is shown relative to B2GP-PLYP-F12-D3(BJ)/VDZ-F12; white = 1, blue = faster, and red = slower.

**3.6. Computational Cost Considerations.** It is of interest to compare the relative computational cost of conventional B2GP-PLYP-D3(BJ) and B2GP-PLYP-F12-D3(BJ) procedures. Each of these timing evaluation jobs was run on otherwise empty nodes with identical hardware (Intel Haswell 2.4 GHz with 256 GB RAM and a 3.6TB SSD RAID array). These jobs were run serially, on a single core, in order to eliminate differences in parallelization as a confounding factor. Timing data relative to VDZ-F12 are reported in Table 6 for the six linear n-alkane dimers in ADIM6,[6,97] (ethane)$_2$ through (n-heptane)$_2$, plus additionally (n-octane)$_2$, (n-nonane)$_2$, (n-decane)$_2$, and (n-dodecane)$_2$ with the structures obtained by manually inserting additional CH$_2$ groups (because we needed them only for timing purposes). As we have seen above, even the VDZ-F12 basis set yields results of a quality comparable to V{Q,5}Z$^m$. In view of the fact that no CP correction is being applied, one would definitely want to use haV{Q,5}Z for this subset rather than V{Q,5}Z. The sum of timings for both calculations involved in the extrapolation ranges from 8 times longer than VDZ-F12 for propane dimer to twice as long for n-dodecane dimer, with the ratio increasingly less favorable to VDZ-F12 as the chain lengths increase. For AV{T,Q}Z, one goes from about 4 times to about the same time, and for def2-{T,Q}ZVPPD that falls from 2 times slower to over twice as fast. In addition, the canonical VDZ-F12 calculations will be increasingly more demanding in mass storage requirements, at least for MOLPRO.

However, in a very recent communication,[148] we have shown that the F12 step can be drastically accelerated by evaluating it in terms of localized pair natural orbitals (i.e., PNO-DHDF-F12) without materially sacrificing accuracy. In addition, it is shown there that the scratch storage requirements are an order of magnitude smaller and that parallelization is more efficient than canonical F12 as well. By way of illustration of what this approximation enables, we offer a PNO-B2GP-PLYP-LMP2/VDZ-F12 calculation on C$_{60}$ (no symmetry): after deleting a diffuse p function that causes insurmountable near-linear dependence issues, it took just 51 min wall clock time on 16 cores of an Ice Lake 2.2 GHz node.

For the discussion at hand here, single-core PNO-DHDF-F12 timing data can be found in the first column of Table 6. For the smallest systems, as expected, the localized approach offers no benefit, but as the chain length increases ever-better speedup is realized, to reach an order of magnitude for n-dodecane dimer. By way of illustration, we applied a power fit to the canonical DHDF-F12 and PNO-DHDF-F12 wall clock times for n-pentane through n-dodecane dimers and found very good fits with $R^2$ = 0.99917 and $R^2$ = 0.99996, respectively: the scaling exponents are 4.57 and 2.60, respectively, showing the scaling advantage of the PNO-F12 approach. Indeed, in ref 148 we show that for still longer chains through n-tetracosane dimer (i.e., n = 24), scaling with n decreases further toward linearity.

It is clear that PNO-DHDF-F12/VDZ-F12 offers a more economical alternative than any of the basis set extrapolations that would yield comparable, or even somewhat inferior, accuracy. We therefore do believe that the DH-F12 approach, especially in its localized orbital form, compares favorably in both accuracy and efficiency with large basis set B2GP-PLYP-D3(BJ).

**3.7. Spin-component-scaled Double Hybrids.** We will now evaluate GMTKN55 performance for the more recent and accurate revDSD-PBEP86-D3(BJ) functional[8] with and without explicit correlation. Table 7 presents statistical analysis for conventional revDSD-PBEP86-D4 and explicitly correlated revDSD-PBEP86-F12-D4 calculations. Using the V$nZ^m$ basis set in conjunction with conventional revDSD-PBEP86-D4 results in WTMAD2 values of 2.236 and 2.104 kcal/mol, respectively, for VQZ$^m$ and V5Z$^m$ basis sets. Leaving out diffuse functions altogether—including in the anionic subsets such as the G21EA electron affinities and the hydroxide clusters in WATER27— WTMAD2 unacceptably increases by 0.6 kcal/mol from VQZ* to VQZ and by 0.4 kcal/mol from V5Z* to V5Z. G21EA alone accounts for 0.176 (VQZ) and 0.077 (V5Z) kcal/mol, respectively.

Finally, the V{Q,5}Z$^m$ pair yields a WTMAD2 of 2.233 kcal/mol. Clearly, in the F12 calculations, WTMAD2 converges spectacularly faster with respect to the basis set size, with even VDZ-F12* reaching statistics comparable to V5Z* in the non-F12 approach. VDZ-F12* and VTZ-F12* yield WTMAD2 values which are 2.233 and 2.218 kcal/mol above the reference values; the latter is close to the "basis set limit" goal as WTMAD2 of V{D,T}Z-F12* is only 0.006 kcal/mol below VTZ-F12*.

At a reviewer's request, we further explored the basis set convergence of the same spin and opposite spin components of the PT2 term in a double hybrid. It is well established (see, e.g., Kutzelnigg and Morgan[25]) that in the large-$L$ limit, MP2 same-spin correlation energies converge as $L^{-5}$ and their opposite-spin counterparts as $L^{-3}$. Hence, for sufficiently large basis sets, opposite-spin will dominate the convergence behavior and same-spin will be effectively saturated.

Although it stands to reason that this would also be the case for GLPT2 correlation, by way of illustration we show in Figure S1 that this is indeed the case for the same-spin ($E_{2ss}$) and opposite-spin ($E_{2ab}$) components of the B2GP-PLYP atomic correlation energy of neon atom along the $n$ZaP basis set sequence ($n$ = 3–8, with maximum angular momentum $L$ = $n$) of Petersson.[149,150] Hence, for sufficiently large $L$, same-spin and opposite-spin contributions in B2GP-PLYP converge as $L^{-5}$ and $L^{-3}$, respectively, and the latter will completely dominate convergence of the overall correlation energy.

## 4. CONCLUSIONS

We have investigated the basis set convergence of double hybrids in conjunction with explicitly correlated (F12) on a large and chemically diverse GMTKN55 database. We chose B2GP-PLYP-D3(BJ) and revDSD-PBEP86-D3(BJ) as test cases. Two families of basis sets were considered: orbital basis sets as large as aug-cc-pV(5+d)Z and F12 basis sets as large as cc-pVQZ-F12. We found that explicitly correlated double-hybrid calculations with F12 basis converge markedly faster than the conventional double-hybrid calculations with orbital (aug-)cc-pV(5+d)Z or def2 basis sets. In fact, DHDF-F12 calculations with just a cc-pVDZ-F12 basis set are closer to the basis set limit than DHDF/cc-pV(Q+d)Z or def2-QZVPPD and approach DHDF/cc-pV(5+d)Z in quality at about one-third the cost. One significant benefit of DHDF-F12 is reducing BSSE by an order of magnitude over orbital-only DHDF in a similar-sized basis set: this particularly benefits the noncovalent interaction subsets (both intermolecular and conformer). We also found that even for anionic systems, the anion-friendly aug-cc-pVDZ-F12 basis set proved unnecessary and cc-pVDZ-F12 was adequate. Finally, although the application of DH-F12 to larger molecules will eventually face mass storage and I/O bandwidth challenges in a disk-based algorithm, these can be circumvented through localized pair natural orbital approaches,[148] which also reduce CPU time scaling by (in practice) about $n^2$.

Table 7. Statistical Analysis of the Basis Set Convergence in Conventional and Explicitly Correlated revDSD-PBEP86-D4 Calculations for the GMTKN55 Database and Its Categories, Relative to the Reference 6 Reference Data

| | revDSD-PBEP86-D4 | | | | | | revDSD-PBEP86-F12-D4 | | | | | |
|---|---|---|---|---|---|---|---|---|---|---|---|---|
| | WTMAD2 | THERMO | BARRIERS | LARGE | CONF | INTERMOL | | WTMAD2 | THERMO | BARRIERS | LARGE | CONF | INTERMOL |
| VQZ | 3.087 | 0.767 | 0.355 | 0.489 | 0.539 | 0.937 | VDZ-F12 | 2.247 | 0.518 | 0.310 | 0.545 | 0.412 | 0.463 |
| VQZ* | 2.494 | 0.547 | 0.244 | 0.489 | 0.539 | 0.675 | VDZ-F12* | 2.233 | 0.513 | 0.306 | 0.545 | 0.412 | 0.458 |
| VQZ$^m$ | 2.236 | 0.547 | 0.244 | 0.487 | 0.399 | 0.559 | VTZ-F12 | 2.216 | 0.520 | 0.308 | 0.530 | 0.397 | 0.462 |
| V5Z | 2.502 | 0.628 | 0.318 | 0.501 | 0.386 | 0.670 | VTZ-F12* | 2.218 | 0.521 | 0.305 | 0.530 | 0.397 | 0.466 |
| V5Z* | 2.101 | 0.496 | 0.226 | 0.501 | 0.386 | 0.492 | V{D,T}Z-F12 | 2.213 | 0.517 | 0.310 | 0.526 | 0.389 | 0.471 |
| V5Z$^m$ | 2.104 | 0.496 | 0.226 | 0.499 | 0.402 | 0.480 | V{D,T}Z-F12* | 2.213 | 0.520 | 0.307 | 0.526 | 0.389 | 0.471 |
| V{Q,5}Z | 2.563 | 0.600 | 0.313 | 0.503 | 0.451 | 0.696 | | | | | | | |
| V{Q,5}Z* | 2.235 | 0.516 | 0.227 | 0.503 | 0.451 | 0.537 | | | | | | | |
| V{Q,5}Z$^m$ | 2.233 | 0.516 | 0.227 | 0.502 | 0.433 | 0.555 | | | | | | | |

Summing up, explicitly corrected double-hybrid calculations are an economical and accurate alternative if (near-)basis set limit results are required, for example, for benchmarking or parametrizing double-hybrid DFT methods. Implementation in other electronic structure systems of MP2-F12 in a basis of Kohn−Sham orbitals would be a very worthwhile endeavor, especially if said implementation is parsimonious in I/O requirements.

## ■ ASSOCIATED CONTENT

### ⓈI Supporting Information

The Supporting Information is available free of charge at https://pubs.acs.org/doi/10.1021/acs.jctc.2c00426.

- Statistical results of all assessed methods, as well as sample MOLPRO input files for B2GP-PLYP-F12-D3(BJ) and revDSD-PBEP86-F12-D4 and Figure S1 (PDF)
- GMTKN55 statistical summary files for various basis set - functional combinations (ZIP)
- Excel spreadsheet with additional timing information (ZIP)

## ■ AUTHOR INFORMATION

### Corresponding Author

Jan M. L. Martin − *Department of Molecular Chemistry and Materials Science, Weizmann Institute of Science, 7610001 Reḥovot, Israel;* orcid.org/0000-0002-0005-5074; Phone: +972-8-9342533; Email: gershom@weizmann.ac.il; Fax: +972-8-9343029

### Author

Nisha Mehta − *Department of Molecular Chemistry and Materials Science, Weizmann Institute of Science, 7610001 Reḥovot, Israel;* orcid.org/0000-0001-7222-4108

Complete contact information is available at:
https://pubs.acs.org/10.1021/acs.jctc.2c00426

### Notes

The authors declare no competing financial interest.

## ■ ACKNOWLEDGMENTS

Work on this paper was supported in part by the Israel Science Foundation (grant 1969/20), by the Minerva Foundation (grant 2020/05), and by a research grant from the Artificial Intelligence and Smart Materials Research Fund, in memory of Dr. Uriel Arnon, Israel.